\documentclass[12pt]{iopart}
\usepackage{graphicx}
\usepackage{psfig}

\begin{document}
\title[Hadron $R_{CP}$ in the forward and backward rapidities in dAu 
collisions]{Hadron production in the forward and 
backward rapidities in dAu collisions at RHIC }

\author{Ming Xiong Liu for the PHENIX Collaboration
\footnote[1]{For the full PHENIX Collaboration author list and acknowledgment, 
see Appendix ``Collaboration'' of this volume.}
}
\address{Physics Division,
Los Alamos National Laboratory, Los Alamos, NM 87545, USA}

\begin{abstract}
We have developed new techniques to detect hadrons with the PHENIX muon 
spectrometers. This allows us to study the centrality dependent 
nuclear modification factor $R_{CP}$ with 
high $p_{T}$ hadrons in both forward (d direction) and backward (Au direction) pseudo-rapidities, 
$1 < |\eta| < 2$, in d-Au collisions at $\sqrt{s_{NN}} = 200 GeV$. 
Preliminary results show a suppression (enhancement) of high $p_{T}$ hadron production
in central $0-20\%$ dAu collisions  
relative to the peripheral one ($60-88\%$ in centrality) at forward (backward) rapidity.

\end{abstract}




\section{Introduction}
The PHENIX experiment has collected $2.74 nb^{-1}$ of dAu collisions from the RHIC 2003 run period.  
This provides the first opportunity to study the cold nuclear medium effects on 
particle production in dAu collisions at $\sqrt{s_{NN}} = 200 GeV$. 
The observed suppression of high $p_{T}$ particles  
in central AuAu collisions and enhancement in dAu collisions at central rapidity at RHIC 
showed the nuclear medium  plays an important role in particle production.\cite{RCP_RHIC} 
In this paper, we extend our measurement of $R_{CP}$, defined in eq. (1), in dAu collisions from 
central pseudo-rapidity $\eta \sim 0$ to the forward and backward regions, 
$1 < |\eta| < 2$. 

\begin{equation}
R_{CP}(p_{T},\eta) = \frac{ \frac{1}{<N_{coll}>}\frac{d^2N}{dp_{T}d\eta}(p_{T},\eta)^{(central)} }
                       { \frac{1}{<N_{coll}>}\frac{d^2N}{dp_{T}d\eta}(p_{T},\eta)^{(peripheral)} }
\end{equation}
where $<N_{coll}>$ is the average number of binary collisions of a given centrality.
For High $p_{T}$ particles, the production cross section is normally given by pQCD,

\begin{equation}
 d\sigma \propto f^{A}(x_{1})\otimes f^{B}(x_{2})\otimes 
D_{h}(z)\otimes d\hat{\sigma}_{x_1 + x_2} 
\end{equation}
where $f^{i}(x)$ is the parton distribution function of the incoming beam particle $i=A,B$,
 $D_{h}(z)$ is the fragmentation function and $d\hat{\sigma}$
is the partonic cross section. 

Nuclear modifications of the initial parton distributions, 
such as (anti)shadowing\cite{SHADOWING}, and variations in parton energy loss 
and multiple scattering could lead to a change in $R_{CP}$. 
It is also important to note that particles produced in dAu collisions 
at large forward (or backward) rapidity are from partons with small (or large) $x$
in Au nuclei, $x = \frac{M_T}{\sqrt{s}}e^{-y}$, with $M_T$ the mass scale of 
the partonic process. 
Without nuclear effects, the number of particles produced in 
hard scatterings would scale with the number of binary collisions 
and $R_{CP}(p_{T},y)$ would be unity, independent of the rapidity and centrality.
Thus $R_{CP}$ provides a good experimental tool to study non-trivial 
nuclear effects in heavy ion collisions. 
Our current understanding of nuclear effects at large rapidity 
in high energy heavy ion collisions is very limited both experimentally and theoretically. 
Several existing models give quite 
different predictions\cite{RCP_TH} on particle 
production at large rapidity - from strong enhancement to strong suppression.

Besides the difference in the initial state $x$ values, it is also 
important to realize that in d-Au collisions at RHIC, particles detected 
in the forward and backward 
directions by the PHENIX muon detector in the CM frame (LAB frame at RHIC) 
have very different kinematics 
in the Au rest frame (from $10^{0}$GeV, backward, to $10^{3}$GeV, forward) 
thus the interactions of the produced particles
with the (cold) Au nuclear medium, right after the hard scattering, 
could be very different in these two kinematic regions\cite{HADRONIZATION}. 
Understanding such effects is important to allow disentanglement of initial 
and final state effects in AuAu collisions.

\section{Hadron measurement with the  PHENIX Muon Spectrometers}
The two PHENIX muon spectrometers cover the pseudo-rapidity ranges $1.2 < \eta < 2.4$ 
and $-2.2 < \eta < -1.2$ with excellent momentum resolution and muon identification\cite{CDR}. 
In front of each muon spectrometer, there is a $5\lambda_I$ (nuclear interaction 
length) thick  nose cone absorber that is about $40$cm away 
from the central collision point. 
The excellent capability for muon measurement has already been  
demonstrated in recently published $J/\psi \rightarrow \mu^{+} \mu^{-}$  
measurements\cite{PHENIX_JPSI}. Here we discuss the novel techniques we 
developed recently to extend the capability of the muon spectrometers to include
hadron measurements at large rapidity. 

\begin{figure}
  \begin{minipage}{7.5cm}
    \hspace{2.0cm}\includegraphics[width=6.0cm,height=4.0cm]{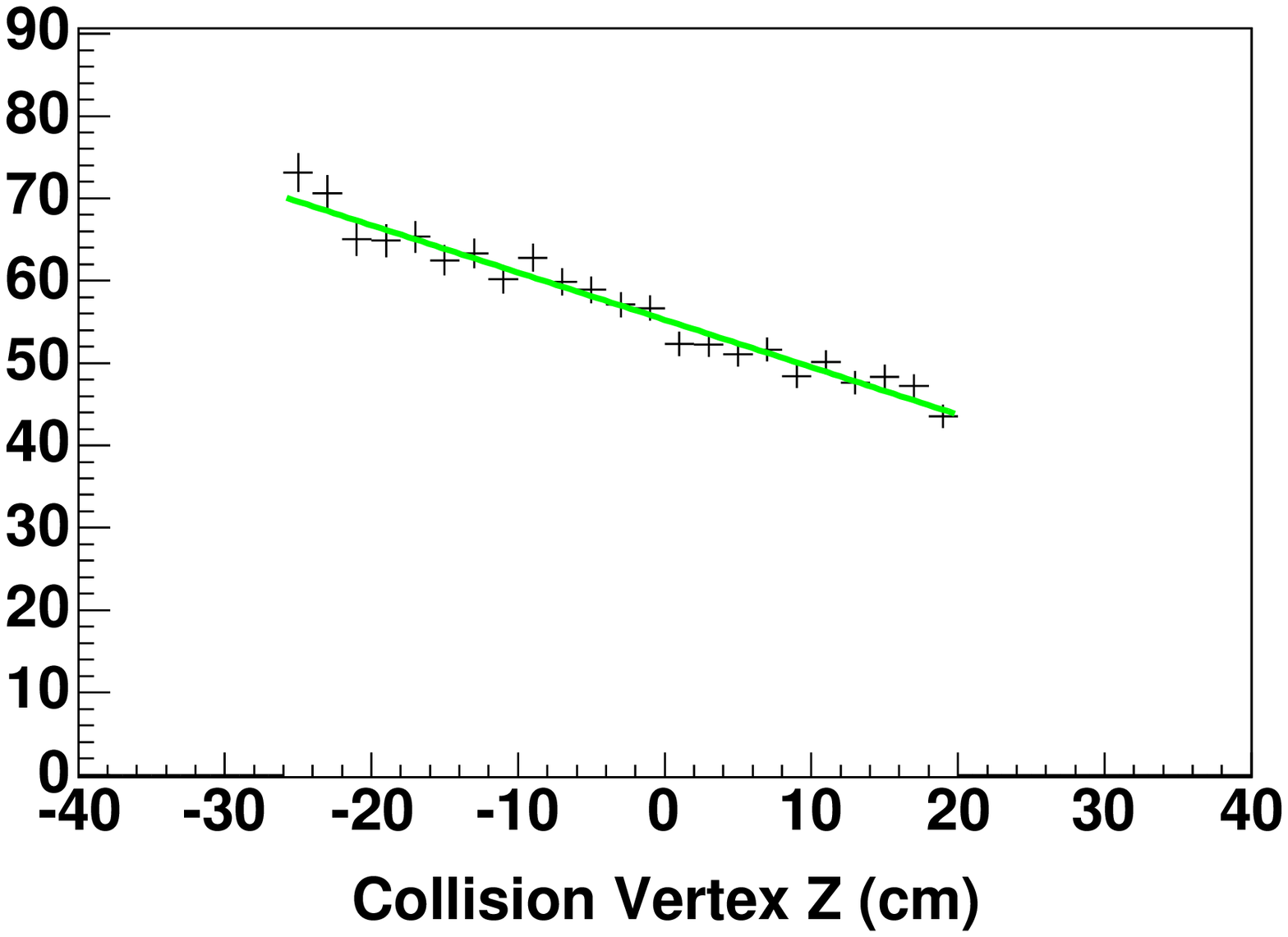}
    \caption{Normalized collision vertex distribution from events with forward muons($p_Z > 0$).}
    \label{VTX-dis}
  \end{minipage}
  \begin{minipage}{7.5cm}
    \hspace{2.0cm}\includegraphics[width=6.5cm,height=4.0cm]{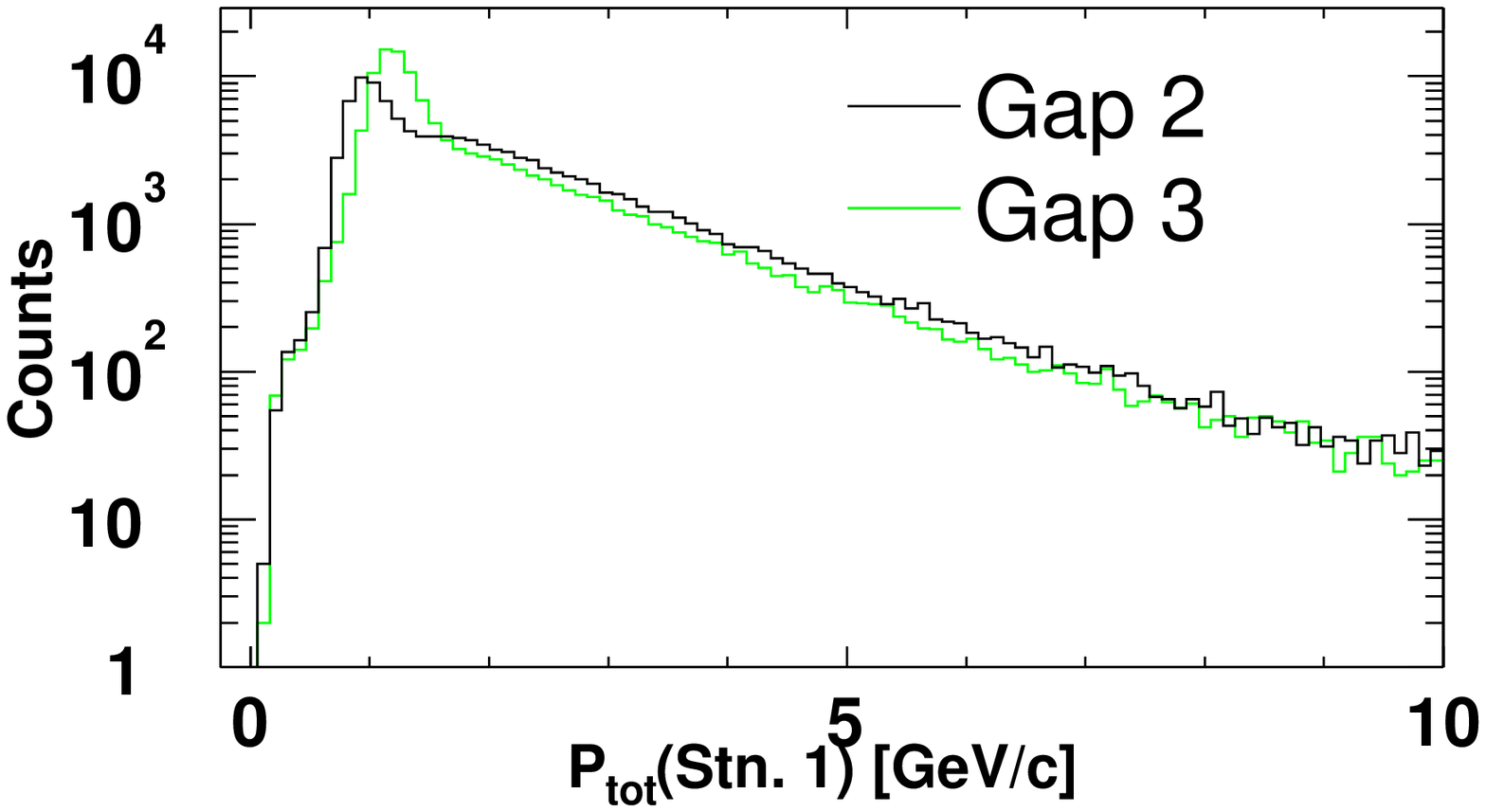}
    \caption{\label{Hadron-P-dis}Momentum distributions 
      of tracks stopped at various muID absorber layers. }
  \end{minipage}
\end{figure}

A hadron can decay into a muon before the nose cone absorber,
the muon is then measured by the muon spectrometer. 
This method is widely used in high energy experiments to measure 
heavy flavor production through their (semi)leptonic decays. 
We extend this method to measure light meson decays, 
such as $\pi^{\pm}\rightarrow \mu^{\pm} + \nu$. 
In PHENIX, due to the finite distance from the collision vertex to 
the nose cone absorber, charged pions and kaons from the collisions
have a chance to decay before they reach the absorber, with  
the decay probability $P_{decay}$ given by,

\begin{equation}
 P_{decay}(p,L) = 1 - \exp(-\frac{L \cdot m}{\tau \cdot p}) \simeq  \frac{L \cdot m}{\tau \cdot p}
\end{equation}
where $L$ is the distance to the absorber; $p$, $m$ and $\tau$ are the momentum,
mass and proper lifetime of the particle. Thus, collisions that occurred far 
from the absorber will be more likely to have muons from light meson decays than 
those that happened close to the absorber.
 \Fref{VTX-dis} shows the normalized collision vertex distribution from events with forward muons 
in dAu collisions. The large slope indicates a significant fraction of muons are  
from pion and kaon decays. By studying the event collision vertex distribution, 
muons from light meson decays are extracted statistically. 
For heavy hadrons, due to their very short proper decay lengths, 
$\exp(-\frac{L \cdot m}{\tau \cdot p}) << 1$, 
the collision vertex dependence is minimal.

In addition to muons from hadron decays, about $1\%$ of hadrons from the collisions can punch through the 
first nose cone absorber, sail through the muon tracking system and finally be
absorbed  inside the muon identification system.   
\Fref{Hadron-P-dis} shows the  momentum 
distributions of tracks stopped in the gap-2 and gap-3 muID absorber 
layers.
The sharp peaks near $P_{tot} \sim 1.5$GeV are due to stopped low energy muons, as they lose 
all of their energy by ionization in the absorber.  
In the high momentum region, $P_{tot} > 2$GeV, most of the tracks are stopped hadrons.
A detailed MC simulation shows that a momentum cut $P_{tot} > 2$GeV can 
reject most of the soft muons and yield about $97\%$ pure hadrons in the sample. 


\section{Results}
 
Hadrons measured with the methods described above are used 
to study the centrality and rapidity dependence of 
$R_{CP}$.  In the first case, most of the muons are from 
an about equal mixture of charged pion and kaon decays; in the second case, 
the tagged hadrons are from a mixture of mesons and baryons, with relative fractions modulated by
their nuclear interactions with the nose cone absorbers. Minimum $p_T > 1.5$GeV 
and $p_T > 1.0$GeV cuts are used in the above analyses to reject soft processes, respectively.
Figures 3 and 4 show the centrality dependent $R_{CP}$  
measured with muons from light meson decays and punch-through hadrons. 
The reference peripheral bin is from $60-88\%$ centrality collisions. $R_{CP}$ results are shown in 
$0-20\%$, $20-40\%$  and $40-60\%$ centrality bins.  
The dominant errors are from statistics and the uncertainties in the determination of the
number of binary collisions $<N_{coll}>$ in a given centrality bin, 
which is shown as a block error bar in each plot at $R_{CP}=1$. 
We have observed a suppression in very central collisions relative 
to the peripheral one in the forward rapidity, 
and an enhancement in the backward rapidity. 
For a typical $P_T \sim 1.5$GeV hadron, the $x$ value probed in Au at the very forward 
rapidity $y \sim 2$ is estimated 
as $x_{Au} \sim \frac{2\times p_T}{\sqrt{s_{NN}}} e^{-y} = 2\times 10^{-3}$; 
at the very backward rapidity $y \sim -2$, 
$x_{Au} \sim \frac{2\times p_T}{\sqrt{s_{NN}}} e^{-y} = 1\times 10^{-1}$.

\begin{figure}
  \begin{minipage}{7.5cm}
    \hspace{0.0cm}\includegraphics[width=8.5cm,height=6.3cm]{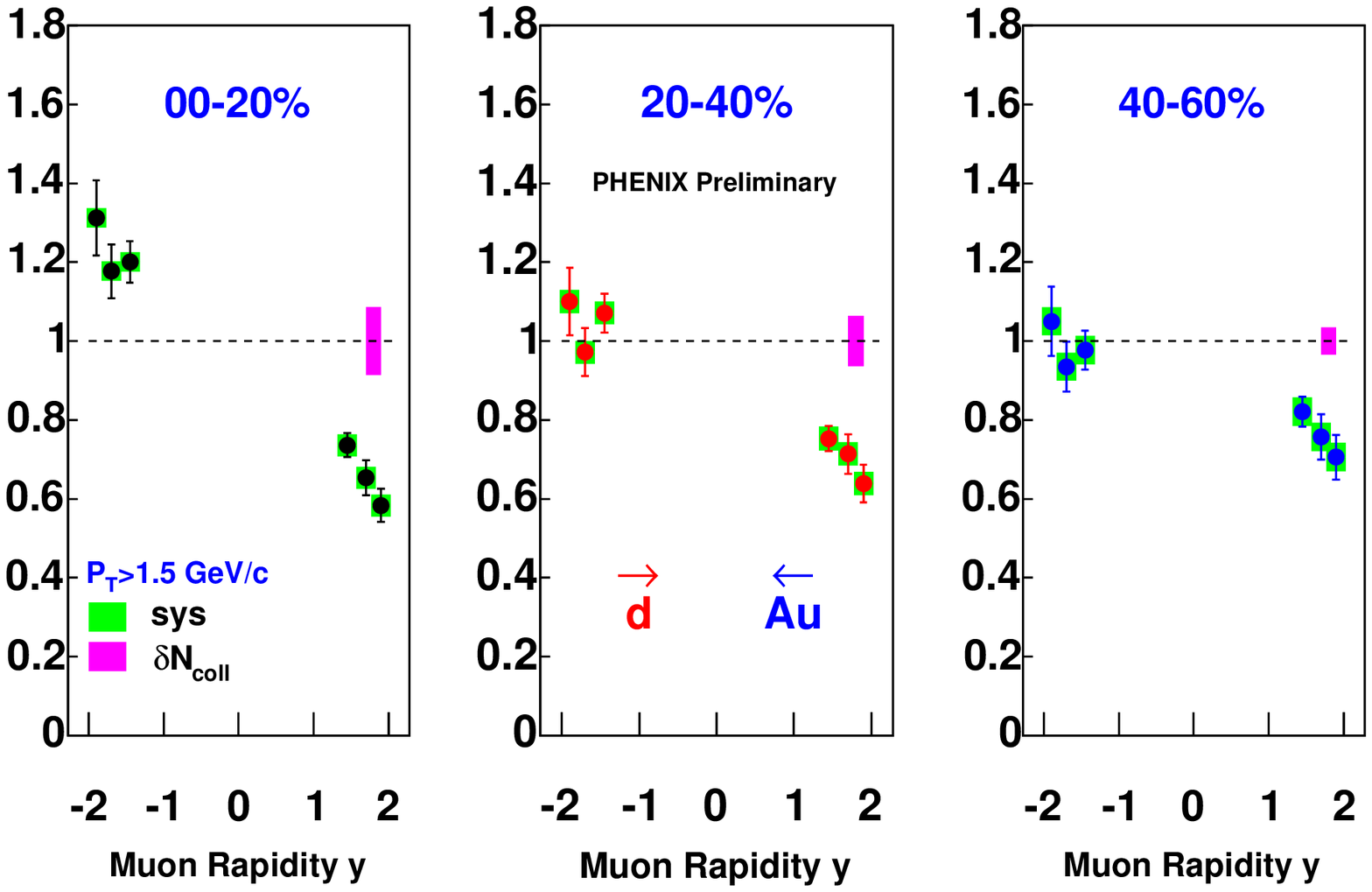}
    \caption{$R_{CP}$ measured with muons from light meson decays.}
    \label{Rcp-muon}
  \end{minipage}
  \begin{minipage}{7.5cm}
    \hspace{1.0cm}\includegraphics[width=7.0cm,height=6.0cm]{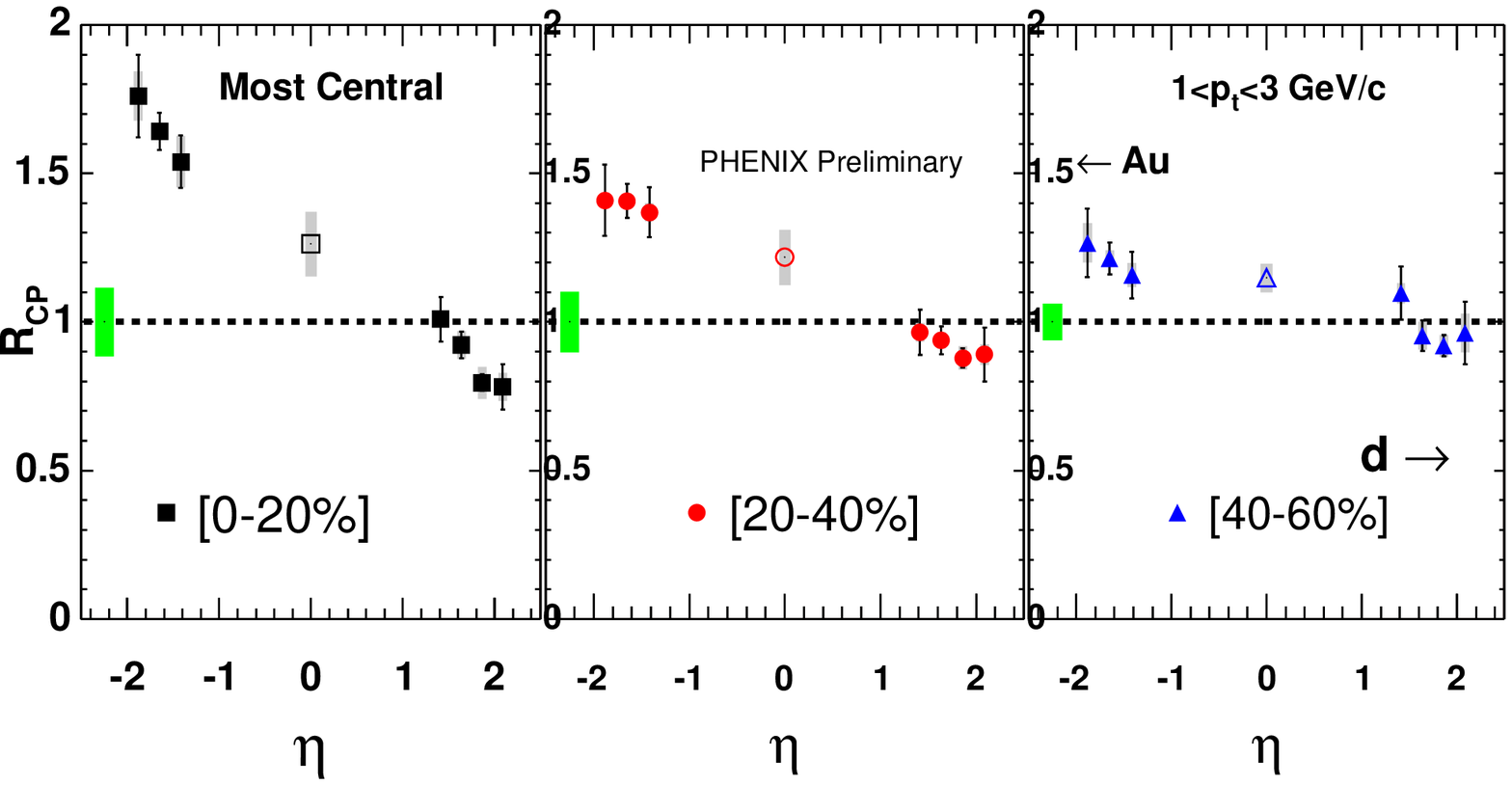}
    \caption{$R_{CP}$ from punch-through hadrons.}
    \label{hadron_rcp}
  \end{minipage}
\end{figure}

\section{Summary and Outlook}

We have presented $R_{CP}$ measurements with hadrons in the forward and backward 
rapidities in dAu collisions at RHIC. BRAHMS experiment 
has also reported similar results at forward rapidity in this conference.\cite{BRAHMS}
The preliminary results seem qualitatively consistent with parton shadowing/saturation picture 
in small $x$ and antishadowing/Cronin effect in large $x$ inside the Au nucleus. 
It will be very interesting to see what happens to Drell-Yan and open charm $R_{CP}$ 
as we have observed a very similar effect in $J/\psi$ production in d-Au collisions.\cite{RCP_JPSI} 
In addition to $R_{CP}$, measurements of $R_{dA}$ (which is normalized 
with pp data rather than peripheral dAu data) will be important to completely 
understand the absolute nuclear effects,   such as suppression and enhancement, since
the most peripheral $60-88\%$ bin used to calculate $R_{CP}$ in this 
analysis could  still be significantly 
different from a minimum bias $pp$ reference. 

 
\section*{References}

\end{document}